\pdfoutput=1

\documentclass[12pt,preprint]{aastex}

\usepackage{graphicx}
\usepackage{natbib}
\usepackage{amsmath}
\usepackage{url}
\usepackage{bm}
\bibpunct{(}{)}{;}{a}{}{,}

\newcommand{\sect}[1]{Sect.~\ref{#1}}
\newcommand{\fig}[1]{Fig.~\ref{#1}}
\newcommand{\eq}[1]{Eq.~\ref{#1}}

\newcommand{\wthreej}[6]{\left( \begin{array}{ccc} #1 & #2 & #3 \\ #4 & #5 & #6 \end{array} \right)}
\newcommand{\gn}{Gaussian}
\newcommand{\ngn}{non--Gaussian}
\newcommand{\ngy}{non--Gaussianity}

\newcommand{\LL}{{\mathrm{L}}}
\newcommand{\NL}{{\mathrm{NL}}}
\newcommand{\fnl}{$f_{\NL}$}

\begin{document}

\title{Improved simulation of non--Gaussian temperature and
  polarization CMB maps}
\shorttitle{Simulation of non--Gaussian CMB maps}

\author{Franz Elsner\altaffilmark{1,*}, Benjamin
D. Wandelt\altaffilmark{2,3,4}}

\altaffiltext{1}{Max--Planck--Institut f\"ur Astrophysik,
Karl--Schwarzschild--Stra\ss e 1, D--85748 Garching, Germany}
\altaffiltext{2}{Department of Physics, University of Illinois at
Urbana-Champaign, 1110 W.~Green Street, Urbana, IL 61801, USA}
\altaffiltext{3}{Department of Astronomy, University of Illinois at
Urbana-Champaign, 1002 W.~Green Street, Urbana, IL 61801, USA}
\altaffiltext{4}{California Institute of Technology, MC 130-33,
Pasadena, CA 91125, USA}
\altaffiltext{*}{Send offprint requests to: felsner@mpa-garching.mpg.de}

\begin{abstract}
  We describe an algorithm to generate temperature and polarization
  maps of the cosmic microwave background radiation containing
  non--Gaussianity of arbitrary local type. We apply an optimized
  quadrature scheme that allows us to predict and control integration
  accuracy, speed up the calculations, and reduce memory consumption
  by an order of magnitude. We generate 1000 non--Gaussian CMB
  temperature and polarization maps up to a multipole moment of
  $\ell_{max} = 1024$. We validate the method and code using the power
  spectrum and the fast cubic (bispectrum) estimator and find
  consistent results. The simulations are provided to the
  community\footnote{Available at
    \url{http://planck.mpa-garching.mpg.de/cmb/fnl-simulations}}.
\end{abstract}


\section{Introduction}
\label{sec:intro}

The simplest models of inflation predict almost perfectly \gn\
primordial fluctuations, generated by a single scalar quantum field in
ground state
\citep{1981PhRvD..23..347G,1983PhRvD..28..679B,1992PhR...215..203M},
but a large number of alternative scenarios can easily be
constructed. To test competing inflationary models, measurements of
statistical properties of the cosmic microwave background (CMB)
radiation have turned out to be of particular importance. Combined
with constraints on the scalar spectral index $n_s$ and the search for
gravitational waves imprinted on the polarization signature, the test
for \ngy\ is a fundamental means to probe the physical processes of
inflation.

Among all inflationary models predicting significant levels of \ngy,
two broad classes can be distinguished. Non--Gaussianity of
\emph{equilateral} type is realized primarily in models with
non--minimal Lagrangian including higher order derivatives
\citep{2004PhRvD..70l3505A, 2005PhRvD..71d3512S, 2005PhRvD..72l3518C,
  2008PhRvD..78f3523L}. Non--Gaussianity of \emph{local} type is
achieved to very good approximation in multi--field inflation
\citep{Moroi:2001ct, Enqvist:2001zp, 2003PhRvD..67b3503L}, or in
cyclic/ekpyrotic universe models \citep{2001PhRvD..64l3522K,
  2002PhRvD..65l6003S,2008PhRvD..77f3533L}.

Concentrating on \emph{local} \ngy, we parameterize the primordial
curvature perturbations, $\Phi$, by introducing an additional
quadratic dependence on a purely \gn\ auxiliary field $\Phi_{\LL}$,
that is local in real space, of the form \citep{2000MNRAS.313..141V,
  2001PhRvD..63f3002K}
\begin{equation}
  \label{eq:phidef}
  \Phi(r)=\Phi_{\LL}(r) + f_{\NL} \, \Phi_{\NL}(r) \, ,
\end{equation}
where $\Phi_{\NL}(r)$ is defined as
\begin{equation}
  \label{eq:phinldef}
  \Phi_{\NL}(r) = \Phi^2_{\LL}(r)-\langle \Phi^2_{\LL}(r) \rangle \, ,
\end{equation}
and \fnl\ is the dimensionless measure of the amplitude of
\ngy. Primordial \ngy\ in the curvature perturbations $\Phi$ will be
encoded in the CMB signal.

Simulations of maps containing \ngy\ of local type have been
extensively used in the context of WMAP data analysis. They play a
crucial role in probing the sensitivity of analysis tools and provide
the opportunity to quantitatively estimate the contribution of
secondary anisotropies or instrumental effects to the measured level
of \ngy\ in experimental data. With the advent of Planck, probing the
\ngn\ contribution within the CMB radiation even more accurately, the
requirements for high resolution, high accuracy simulations of \ngn\
CMB temperature and polarization maps will further increase.

To meet the demand for simulated \ngn\ maps, several different
approaches have been taken. First simulations of temperature maps with
primordial \ngy\ of local type have generated the underlying
primordial perturbation in Fourier space
\citep{2003ApJS..148..119K}. This approach is computationally very
demanding while it is difficult to preserve numerical accuracy. A
different method has been proposed in \cite{2003ApJ...597...57L,
  2007PhRvD..76j5016L}, where the authors work with \lq filter'
functions to introduce the proper spatial correlations of the
primordial potential. Recently, a fast, specifically tailored
algorithm for the weakly \ngn\ regime has been introduced by
\cite{2006astro.ph.12571S}, that focuses on simulating maps with a
given three--point function. While it is not restricted to \ngy\ of
local type, higher order correlations are not guaranteed to match the
model.

The algorithm presented here was closely inspired by the work of
\cite{2003ApJ...597...57L}. We focus on an enhancement of their
algorithm in view of its numerical efficiency. Our idea is to
precompute quadrature nodes and weights; this is similar in spirit to
\cite{2006astro.ph.12571S}, but aims at assuring accurately simulated
maps to all correlation orders, rather than focusing exclusively on
the three--point function.

This paper is organized as follows. In \sect{sec:basics}, we present a
new approach to simulate \ngn\ temperature and polarization maps. An
optimization scheme is provided in \sect{sec:code} that allows for an
increase in computational efficiency. We then apply the fast estimator
to simulated CMB maps to check our results for consistency
(\sect{sec:analysis}). Finally, we summarize our findings in
\sect{sec:summary}.

Throughout the paper we assume the following {\emph WMAP5+BAO+SN}
cosmological parameters\footnote{Obtained from
  \url{http://lambda.gsfc.nasa.gov/product/map/dr3/parameters.cfm}}:
$\Omega_{\Lambda} = 0.721$, $\Omega_{c} \, h^2 = 0.1143$, $\Omega_{b}
\, h^2 = 0.02256$, $\Delta^2_{\mathcal{R}}(0.002 \ Mpc^{-1}) = 2.457 \cdot
10^{-9}$, $h = 0.701$, $n_{s} = 0.96$, and $\tau = 0.084$.

\section{Simulation of non--Gaussian CMB maps}
\label{sec:basics}

We describe a new, direct method to simulate \ngn\ CMB temperature and
polarization maps below. Our objective is to generate a set of linear
and non--linear spherical harmonic coefficients that are valid
realizations of temperature and polarization fluctuations, $\{a_{\LL
  \ \ell m}, \ a_{\NL \ \ell m}\}$, for a given cosmological
model. A map with any desired level of \ngy, \fnl, can then be
realized by linear combination,
\begin{equation}
a_{\ell m} = a_{\LL \ \ell m} + f_{\NL} \cdot a_{\NL \ \ell m} \, .
\end{equation}

The expansion coefficients $a_{\ell m}$ of the CMB temperature and
polarization anisotropies in harmonic space are related to the
primordial fluctuations $\Phi_{\ell m}(k)$ via the equation
\citep{2003ApJS..148..119K}
\begin{equation}
  \label{eq:phi2alm_k}
  a^{i}_{\ell m}=\frac{(-\imath)^{\ell}}{2\pi^2} \int dk \, k^2 \ 
  \Phi_{\ell m}(k) \ g^{i}_{\ell}(k) \, .
\end{equation}
Here, $g^{i}_{\ell}(k)$ is the transfer function of temperature ($i =
T$) or polarization ($i = E$) in momentum space. Analogously, we can
define an equivalent equation as a function of comoving distance,
\begin{equation}
  \label{eq:phi2alm_r}
  a^{i}_{\ell m} = \int dr \, r^2 \ \Phi_{\ell m}(r) \ \alpha_\ell^i(r) \, ,
\end{equation}
where we have used the real space transfer function according to
\begin{equation}
  \label{eq:rs_tf}
  \alpha_\ell^i(r) = \frac{2}{\pi} \int dk \, k^2 \ g_\ell^i(k) \ j_\ell(kr) \, ,
\end{equation}
where $j_\ell(kr)$ denotes the spherical Bessel function of order
$\ell$.

We can now outline our recipe for simulating \ngn\ CMB maps as
following: (i) Generate the multipole moments of a purely \gn\
gravitational potential $\Phi_{\LL \ \ell m}(r)$ as a function of
conformal distance. (ii) Compute the spherical harmonic transform to
derive the corresponding expression in pixel space,
$\Phi_{\LL}(r)$. (iii) Square it and subtract the variance according
to \eq{eq:phinldef} to get the \ngn\ potential $\Phi_{\NL}(r)$. (iv)
Inverse transform to spherical harmonic space to obtain $\Phi_{\NL \
  \ell m}(r)$. (v) Solve the radial integral \eq{eq:phi2alm_r} for
$\Phi_{\LL \ \ell m}(r)$ and $\Phi_{\NL \ \ell m}(r)$ separately
to compute $\{a^T_{\LL \ \ell m}, \ a^E_{\LL \ \ell m}; \
a^T_{\NL \ \ell m}, \ a^E_{\NL \ \ell m} \}$.

One difficulty in this approach is that we have to take into account
the radial correlation of the gravitational potential in step (i). Its
covariance matrix is determined by the primordial power spectrum
predicted by inflation, $\mathcal{P}(k)$, and is given by
\citep{2003ApJ...597...57L}
\begin{equation}
  \label{eq:cov_phi}
  \left\langle \Phi_{\LL \ \ell_1 m_1}(r_1) \, \Phi^{*}_{\LL \ \ell_2 m_2}(r_2)
  \right\rangle = 4 \pi \ \delta_{\ell_2}^{\ell_1}\ \delta_{m_2}^{m_1}
  \ \int dk \ \frac{\Delta^2_{\mathcal{R}}(k)}{k} \ j_{\ell_1}(kr_1) \ j_{\ell_2}(kr_2) \, ,
\end{equation}
where we have replaced $\mathcal{P}(k)$ by
\begin{equation} 
  \Delta^2_{\mathcal{R}}(k) = \frac{k^3}{2 \pi^2} \cdot {\mathcal{P}}(k) \, ,
\end{equation}
that is constant for vanishing spectral tilt ($n_s = 1$). The
covariance matrix will be denoted by $P_{\Phi \ \ell}(r_1,r_2)$ in
what follows. To draw a random realization of the linear gravitational
potential at distances $\bm{r} = (r_1,r_2,\dots,r_n)$, we calculate
\begin{equation} 
  \Phi_{\LL \ \ell m}(\bm{r}) = P^{1/2}_{\Phi \ \ell} \cdot \bm{g} \, ,
\end{equation}
where $\bm{g}$ is a vector of independent complex \gn\ random
variables with zero mean and unit variance.

For this algorithm to run efficiently, we have to reduce the number of
quadrature points in the numerical evaluation of the radial integral
(\eq{eq:phi2alm_r}), to keep the number of computationally
expensive spherical harmonic transformations necessary to generate the
\ngn\ gravitational potential as low as possible. Details of the
implementation together with an optimization scheme will be described
in the next section.

\section{Implementation and Optimization}
\label{sec:code}

\begin{figure*}
  \label{fig:rs_tf}
  \centerline{\resizebox{\hsize}{!}{\includegraphics*{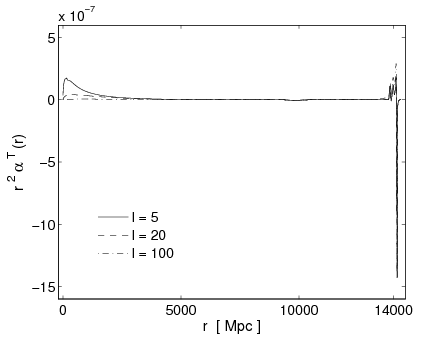} \,
      \includegraphics*{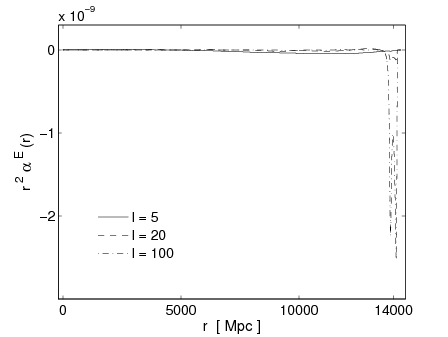}}}
  \caption{Real space transfer functions. We show examples of the real
    space transfer functions of temperature (\emph{left panel}) and
    polarization (\emph{right panel}) for three different multipole
    moments $\ell = 5$, 20, and 100. At low $\ell$, the effect of the
    late time ISW effect is clearly visible. Reionization occurred at
    about $r = 10 \ Gpc$.}
\end{figure*}

To be able to perform the steps outlined in the last section, we first
have to precompute the necessary auxiliary data. This needs to be done
only once for a given set of cosmological parameters. First, we
obtained the transfer functions in momentum space from a modified
version of the latest CAMB software package\footnote{Obtained from
  \url{http://camb.info}} \citep{2000ApJ...538..473L}. We then derived
their equivalent expressions in real space using
\eq{eq:rs_tf}. Examples of temperature and polarization transfer
functions as a function of conformal distance for several multipole
moments are shown in \fig{fig:rs_tf}.

As a next step, we calculate the covariance matrix of the
gravitational potential on a fine grid with $N_{tot} = 400$ shells
from the origin to the present time cosmic horizon
(\eq{eq:cov_phi}). As a start, we resolve the last scattering
surface with uniform spacing using an increment of $\Delta r \approx
3.5 \ Mpc$ and chose a larger interval elsewhere ($\Delta r \approx
100 \ Mpc$). This simple approach will be refined later. Using the
derived quantities, it is now possible to generate $\Phi_{\LL \
  \ell m}(r_{i}), \ \Phi_{\NL \ \ell m}(r_{i})$ and numerically
solve the radial integral \eq{eq:phi2alm_r} to obtain simulated
\ngn\ CMB maps. However, significant improvement in the numerical
evaluation of the integral is achievable by choosing both weights and
quadrature points in an optimal way, as we will show in the following.

\begin{figure*}
  \label{fig:ev_nodes}
  \centerline{\resizebox{\hsize}{!}{\includegraphics*{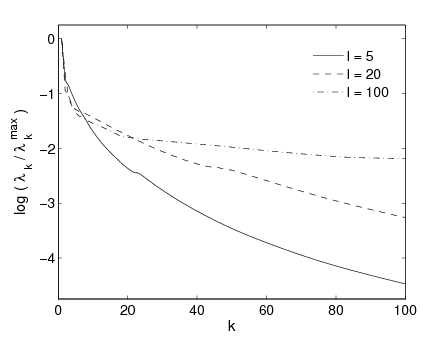} \,
      \includegraphics*{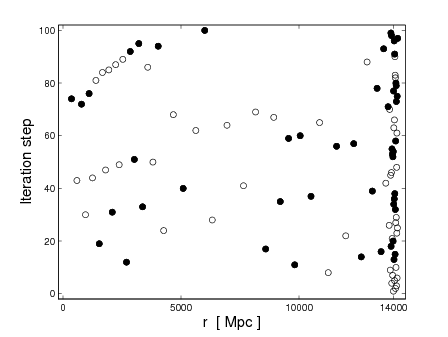}}}
  \caption{Optimization scheme. \emph{Left panel:} We display the
    largest 100 eigenvalues $\lambda_{k}$ of the covariance matrix of
    $\Phi_{\LL \ \ell}(r)$, normalized and in descending order for
    $\ell = 5, \ 20$, and 100. For low multipole moments, the number
    of quadrature points can be reduced most efficiently. \emph{Right
      panel:} The radial positions of the shells included in the first
    100 iteration steps. For illustrative purposes, we interchanged
    open and filled symbols every 10 iterations. The most important
    nodes are included first.}
\end{figure*}

Keeping the multipole moment $(\ell, m)$ fixed for simplicity, we want
to accurately compute the integral
\begin{equation}
  \label{eq:los_int}
  I = \int dr \, r^2 \ \alpha(r) \ \Phi(r) \, .
\end{equation}
This is done in a two step process: We first approximate
\eq{eq:los_int} with a discrete sum over $N_{tot}$ shells. Then, we
try to obtain comparable accuracy with fewer shells $N \ll N_{tot}$
introducing weights,
\begin{equation}
  \hat{I} = \sum^{N}_{i=1} w_{i} \ \Phi(r_{i}) \, ,
\end{equation}
where the gravitational potential is evaluated at the nodes $r_{i}$ and
weighted by the factors $w_{i}$. Now, we can derive the expectation
value of the quadratic error
\begin{equation}
  \label{eq:mean_err}
  \langle (\hat{I} - I)^2 \rangle = \sum^{N_{tot}}_{k=1} \ \lambda_{k} \ \left( \int
  dr \, r^2 \ \alpha(r) \ \phi_{k}(r) \right)^2 - w^{T} P_{\Phi} w \, ,
\end{equation}
where we have introduced the eigenvalues $\lambda_{k}$ and
eigenvectors $\phi_{k}$ of the covariance matrix of the potential on
the fine grid with $N_{tot} = 400$ elements. We show $\lambda_{k}$ for
several multipole moments in the left panel of \fig{fig:ev_nodes}. If
the eigenvalues decrease sufficiently fast, the error is expected to
be low already for a small number of quadrature points $N$. This seems
especially to be true on large angular scales. However, this finding
is partially counterbalanced by the fact that the transfer functions
are significantly different from zero at small radii for low multipole
moments (late ISW effect, reionization), enforcing the inclusion of
additional nodes.

Based on the expression for the expected quadratic error, it is
straightforward to calculate optimal weights by satisfying the
condition $\frac{\partial}{\partial w_i} \langle (\hat{I} - I)^2
\rangle = 0$, which leads to a system of $N$ linear equations,
\begin{equation}
  \sum^{N}_{j=1} P_{\Phi \, i j}(r_{i},r_{j}) \ w_{j} = \sum^{N_{tot}}_{k=1} \
  \lambda_{k} \ \phi_{k}(r_{i}) \ \int dr \, r^2 \ \alpha(r) \
  \phi_{k}(r) \, .
\end{equation}

\begin{figure*}
  \label{fig:error_powers}
  \centerline{\resizebox{\hsize}{!}{\includegraphics*{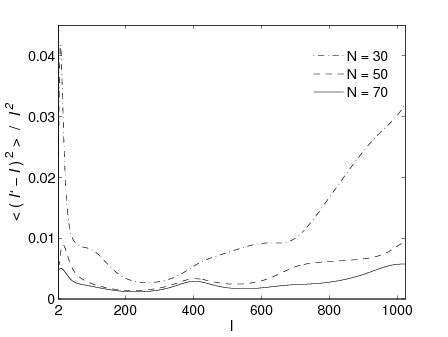} \,
      \includegraphics*{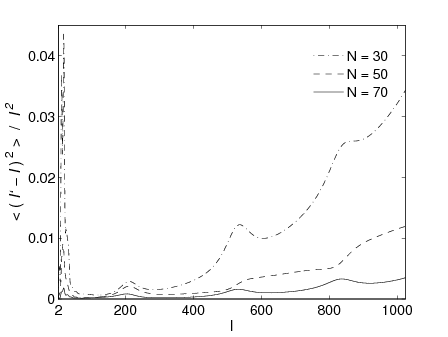}}}
  \caption{Error of integration. We depict the relative mean quadratic
    error introduced by approximating the integral \eq{eq:phi2alm_r}
    by a sum over $N = 30, \ 50$, and 70 elements for temperature
    (\emph{left panel}) and polarization (\emph{right panel}).}
\end{figure*}

Even more important, \eq{eq:mean_err} allows us to formulate a greedy
algorithm to compute optimal quadrature points. We select a subset of
nodes out of the fine radial grid with 400 elements iteratively, in
each step including the point that most efficiently reduces the
remaining error. To simultaneously optimize for temperature and
polarization, we add the expectation values of the two errors with
equal weights. We use the outcome of the procedure to tune the radii
of the input grid with 400 elements. We choose a smaller spacing down
to $\Delta r = 1.2 \ Mpc$ at the last scattering surface, where nodes
were selected with the highest priority, and a larger step size up to
$\Delta r = 140 \ Mpc$ at distances, where the quadrature points were
classified as less important. Then, we repeated the optimization
process a second time. In the right panel of \fig{fig:ev_nodes}, we
visualize the first 100 iterations of the optimization scheme. We
display the expectation value of the relative quadratic error for $N =
30, \ 50$ and $N = 70$ quadrature points in
\fig{fig:error_powers}. The raise in error towards the largest angular
scales is caused by the increasing contribution from late ISW effect
and reionization.

As a last step, we show how to reduce the memory consumption of the
code. Whereas the potentials $\Phi_{\LL \ \ell m}(r)$ and
$\Phi_{\NL \ \ell m}(r)$ can be co--added to the spherical
harmonic coefficients of the CMB map $a^{i}_{\LL \ \ell m}$ and
$a^{i}_{\NL \ \ell m}$ (\eq{eq:phi2alm_r}) one shell after
another, the radial correlation of $\Phi_{\LL}(r)$ forces one to
generate it at all $N$ shells simultaneously, potentially requiring
large amounts of memory. To circumvent this problem, we keep the
random seeds that were used to draw the potential. By means of the
seeds, we are able to easily regenerate the gravitational potential at
any radii $(r_{1}, r_{2}, \dots, r_{N})$. Thus, we only store its real
space representation at the radius that is currently added to the CMB
map, substantially reducing the overall memory consumption of the
algorithm.

Having optimized the simulation algorithm in this way, we generated
$N_{sim} = 1000$ realizations of temperature and polarization CMB
maps. We chose a HEALPix resolution parameter of $n_{side} = 512$ and
a maximum multipole moment of $\ell_{max} = 1024$. We used $N = 70$
quadrature points for evaluation of \eq{eq:phi2alm_r}, although we
stress that this choice is conservative and it is possible to derive
reasonable results with smaller values of $N$.

\begin{figure*}
  \label{fig:cmb_maps}
  \centerline{\resizebox{\hsize}{!}{\includegraphics*{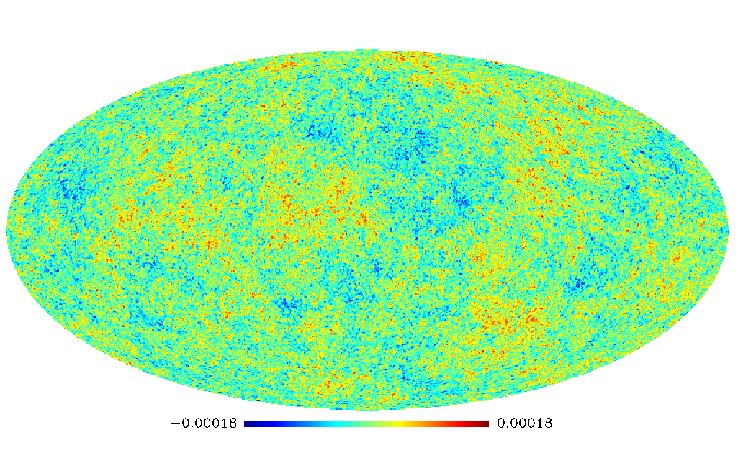} \,
      \includegraphics*{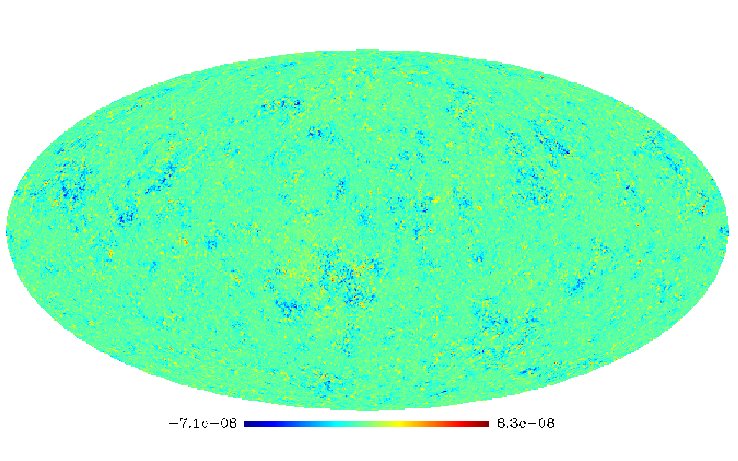}}}
  \centerline{\resizebox{\hsize}{!}{\includegraphics*{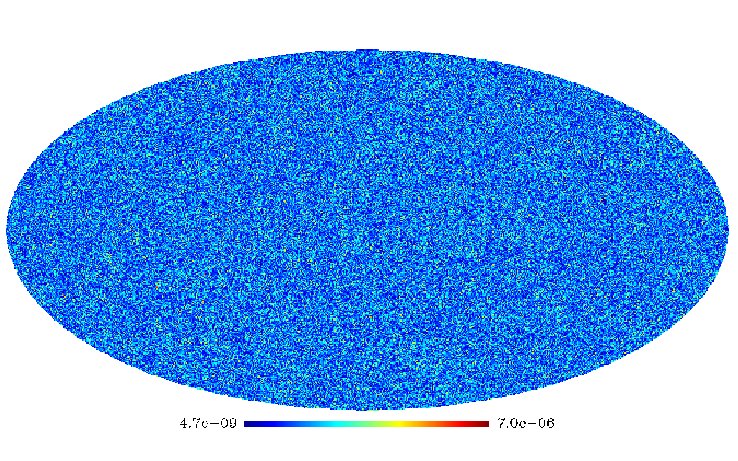} \,
      \includegraphics*{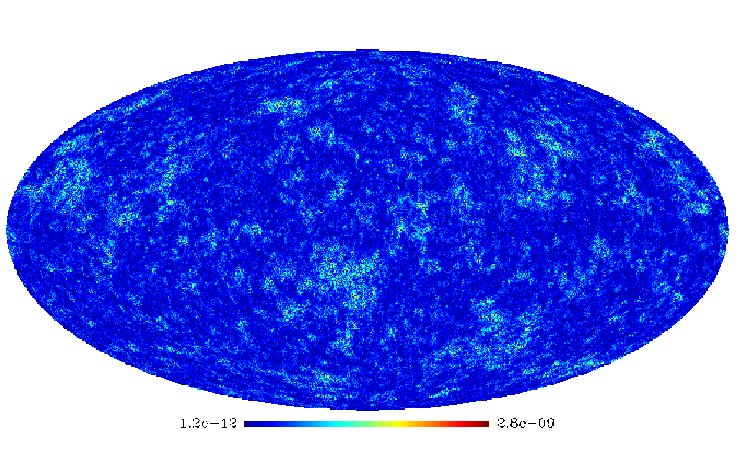}}}
  \caption{Example of simulated \ngn\ map. We show the linear
    (\emph{left column}) and the associated non--linear part
    (\emph{right column}) of a realization of temperature (\emph{first
      row}) and polarization intensity (\emph{second row}) CMB
    data. The polarization intensity is defined as $I = \sqrt{Q^2 +
      U^2}$, where Q and U are the Stokes parameters.}
\end{figure*}

\begin{figure*}
  \label{fig:sim_powers}
  \centerline{\resizebox{\hsize}{!}{\includegraphics*{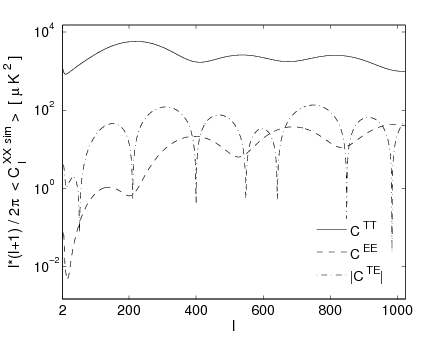} \,
      \includegraphics*{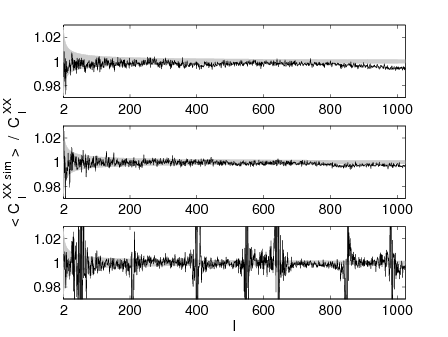}}}
  \caption{Averaged power spectra. \emph{Left panel:} We display the
    power spectra $\mathcal{C}^{TT}, \ \mathcal{C}^{EE}$, and
    $\mathcal{C}^{TE}$ of the linear part of the simulated CMB maps,
    averaged over 1000 simulations. We do not show the input power
    spectra here, as the lines cannot be discerned in this
    view. \emph{Right panels:} The ratio of the power spectra divided
    by their theoretical values for temperature ($XX = TT$,
    \emph{upper sub--panel}), polarization ($XX = EE$, \emph{middle
      sub--panel}), and cross--power spectrum ($XX = TE$, \emph{lower
      sub--panel}). Oscillatory features in the latter are caused by
    roots of the denominator. The grayish area indicates the
    2-$\sigma$ bounds of an ideal simulation code. Sub-percentage,
    systematic deviations for the TT and EE spectra remain but are
    consistent with the precision goal.}
\end{figure*}

With these input parameters, we aim for sub--percentage accuracy of
the final map over the entire range of multipole moments, guided by
the intrinsic precision of the underlying transfer functions, running
CAMB with RECFAST at standard accuracy. An example is displayed in
\fig{fig:cmb_maps}, where we illustrate a realization of temperature
and polarization maps of the linear and non--linear part of the
CMB. We show the averaged power spectra of all simulations along with
a comparison to the theoretical values in \fig{fig:sim_powers}. A
detailed comparison to the expected statistical fluctuations ($\propto
N_{sim}^{-1/2}$) reveals remaining slight systematic deviations for
the TT and EE spectra at high $\ell$ at the level of less than 1\% of
the input power spectrum. If required this error could be further
reduced by adding integration nodes. It takes about 20 minutes to
generate a single map with the given resolution on a single Intel
Xenon processor with a clock rate of 2.33 GHz, requiring only a modest
amount of memory ($\approx 400 \ MB$). The most time consuming part is
the evaluation of the spherical harmonic transforms necessary to
compute the \ngn\ potential.

\begin{figure*}
  \label{fig:phi}
  \centerline{\resizebox{\hsize}{!}{\includegraphics*{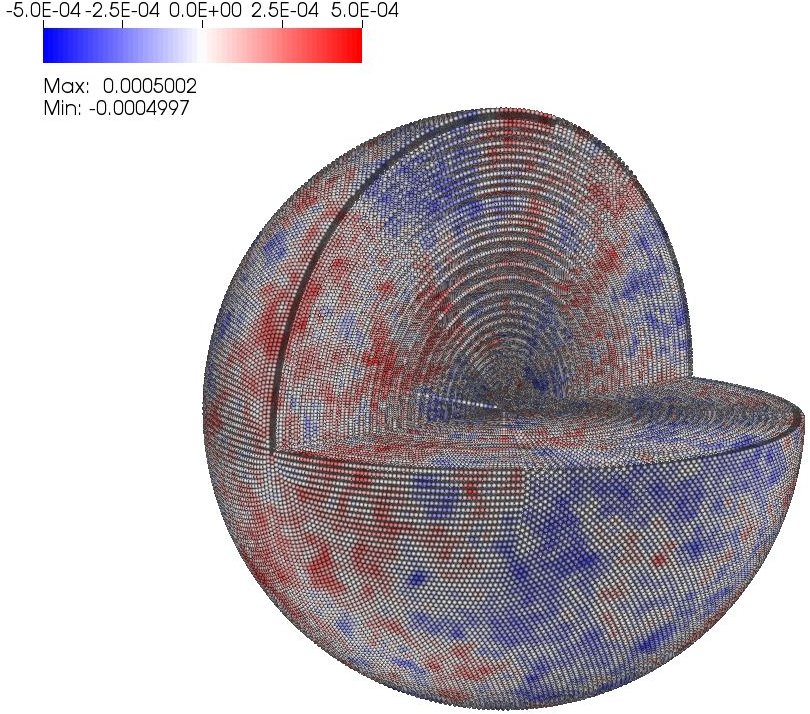} \,
      \includegraphics*{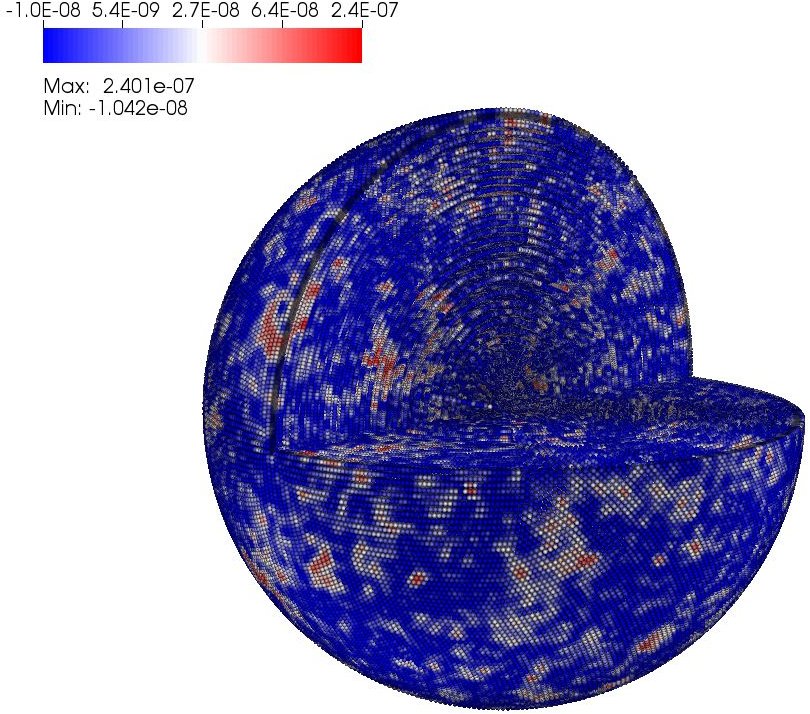}}}
  \caption{Examples for simulated curvature perturbations. \emph{Left
      panel:} We visualize the linear gravitational potentials
    $\Phi_{\LL}(r)$, generated on $N = 70$ shells from the origin
    (\emph{center}) to the last scattering surface (\emph{outermost
      shells}). \emph{Right panel:} The associated \ngn\ potential,
    displayed at nonlinear scale.}
\end{figure*}

The algorithm described here generates valid realizations of
primordial curvature perturbations in real space. This itself is an
interesting quantity and can be used to e.g. test the performance of
reconstruction techniques as we will show in the next section. We
visualize the 3D gravitational potential $\Phi_{\LL}(r)$ and
$\Phi_{\NL}(r)$ in \fig{fig:phi}; long--distance correlations on
large scales are in evidence.

Our simulation algorithm is conceptually very similar to the method
proposed in \cite{2007PhRvD..76j5016L}, where the authors generate the
gravitational potential on 400 shells, requiring 800 spherical
harmonic transforms to calculate a single \ngn\ CMB map, and report a
runtime of 3 hours for $\ell_{max} = 500$. By applying our optimized
quadrature scheme, we have demonstrated that it is possible to reduce
the number of transforms considerably, resulting in an increase of
computational efficiency. Another, albeit more formal difference is
the way the gravitational potential is generated. We use the real
space covariance matrix to draw $\Phi_{\LL}(r)$ directly, whereas
the authors of \cite{2007PhRvD..76j5016L} compute the gravitational
potential by performing an integral over uncorrelated random numbers
weighted by \lq filter' functions.

In \cite{2006astro.ph.12571S}, where the authors focused on a
perturbative reproduction of the correct bispectrum in the regime of
weak \ngy, a runtime of about three minutes is reported to simulate
one \ngn\ CMB temperature map at an angular resolution of $\ell_{max}
= 1000$. Although slower by an order of magnitude, and tuned for local
\ngy, the algorithm presented here is capable of simulating both
temperature and polarization maps (i.e. three maps for the stokes
parameter I, Q, and U) within the same framework and with nearly the
same computational cost compared to temperature alone.  Furthermore,
as recently pointed out by \cite{2009arXiv0905.4732H}, in the case of
local \ngy\ an additional modification of the algorithm of
\cite{2006astro.ph.12571S} is necessary to suppress the power spectrum
of the \ngn\ part of a simulated map, found to be artificially
enhanced by several orders of magnitudes on large angular scales.

Notwithstanding the aforementioned higher computational costs, we
regard our method as useful for the study of local \ngy, because the
simulated maps are well suited to test any kind of estimator,
e.g. based on Minkowsky functionals
\citep{2007ApJS..170..377S,2008MNRAS.389.1439H}, or a wavelet analysis
\citep{2002MNRAS.336...22M,2004ApJ...613...51M}. If a detection of
nonzero \fnl\ is reported, it will be important to confirm the result
with alternative statistical tools, as they are sensitive to different
systematic effects.

In the following section, we apply the KSW estimator
\citep{2005ApJ...634...14K} to our set of simulated maps with known
\ngn\ contribution to test whether the input values for \fnl\ can be
recovered.

\section{Bispectrum Analysis}
\label{sec:analysis}

As we do not aim to describe the fast estimator in detail, we include
a brief summary here and refer the reader to the extensive literature
for further details and a comprehensive discussion
\citep[e.g. in][]{2005ApJ...634...14K, 2006astro.ph.12571S,
  2007JCAP...03..019C, 2007ApJ...664..680Y}.

\begin{figure*}
  \label{fig:wfr_phi}
  \centerline{\resizebox{\hsize}{!}{\includegraphics*{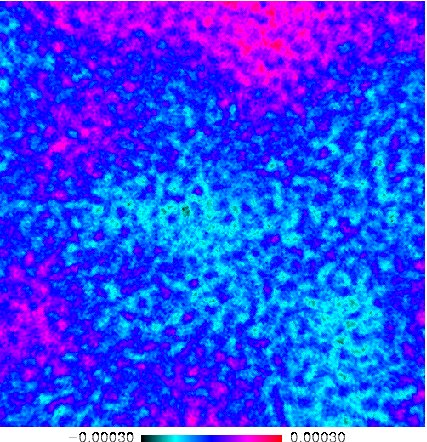} \,
      \includegraphics*{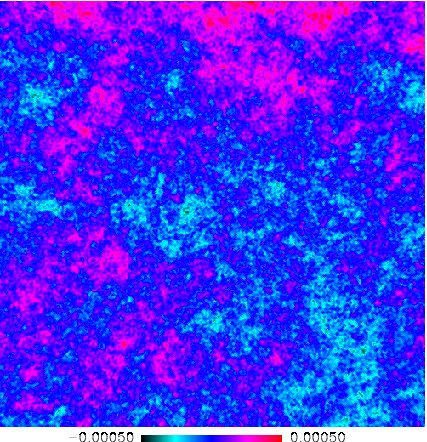} \,
      \includegraphics*{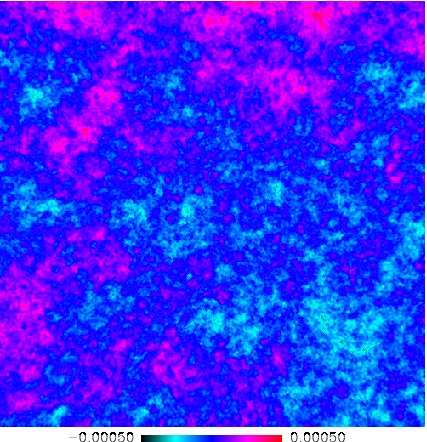}}}
  \caption{Wiener filter reconstruction of the gravitational
    potential. We illustrate the input gravitational potentials
    $\Phi_{\LL}(r)$ at the last scattering surface $r = 14.0 \ Gpc$
    (\emph{middle panel}) used to generate a simulated CMB map and its
    Wiener filter reconstruction based solely on temperature data
    (\emph{left panel}), and based on both, temperature and
    polarization data (\emph{right panel}) of the same map. Each patch
    is $50^{\circ}$ on the side.}
\end{figure*}

To estimate the \ngy\ of a CMB map, one constructs the statistic
$\mathcal{S}_{prim}$ out of a cubic combination of the data,
\begin{equation}
  \mathcal{S}_{prim} = \int dr \, r^2 \ \int d^2\hat{n} \ A(r, \hat{n}) \
  B^2(r, \hat{n}) \, .
\end{equation}
The radial integral runs over two filtered maps,
\begin{align}
 A(r, \hat{n}) &= \sum_{i,j=T,E} \ \sum_{\ell, m}
 (\mathcal{C}^{-1})^{i j}_{\ell} \alpha^i_{\ell}(r) \ a^j_{\ell m} \
 Y_{\ell m}(\hat{n}) \, ,\\
 B(r, \hat{n}) &= \sum_{i,j=T,E} \ \sum_{\ell, m}
 (\mathcal{C}^{-1})^{i j}_{\ell} \beta^i_{\ell}(r) \ a^j_{\ell m} \
 Y_{\ell m}(\hat{n}) \, ,
\end{align}
that are constructed using the auxiliary functions
\begin{align}
  \alpha_\ell^i(r) &= \frac{2}{\pi} \int dk \, k^2 \ g_\ell^i(k) \ j_\ell(kr) \, ,\\
 \beta^i_{\ell}(r) &= 4 \pi \int dk \, \frac{\Delta^2_{\mathcal{R}}(k)}{k} \ 
 g^i_{\ell}(k) \ j_{\ell}(k r) \, ,
\end{align}
and the inverse of the matrix containing the CMB power spectrum
elements,
\begin{equation}
  \mathcal{C}^{-1}_{\ell} = \left( \begin{array}{cc} \mathcal{C}_{\ell}^{TT} &
      \mathcal{C}_{\ell}^{TE} \\ \mathcal{C}_{\ell}^{TE} &
      \mathcal{C}_{\ell}^{EE}\end{array} \right)^{-1} .
\end{equation}
One of these maps, $B(r, \hat{n})$, is exactly the Wiener filter
reconstruction of the underlying gravitational potential
$\Phi(r)$. With the simulation algorithm presented in
\sect{sec:basics}, it is possible to compare the potential used to
synthesize the map with its reconstruction directly. An example is
shown in \fig{fig:wfr_phi}, where we depict the reconstruction of the
potential around last scattering using only temperature, and using
both, temperature and polarization information.

As the estimator $\mathcal{S}_{prim}$ is proportional to the \ngy\
parameter \fnl, we can calculate its expectation value by applying a
suitable normalization,
\begin{equation}
  \label{eq:fast_est}
  f_{\NL} = \left[ \mathop{\sum_{i,j,k,o,p,q}}_{=T,E} \ \sum_{\ell_1
      \le \ell_2 \le \ell_3} \ \frac{1}{\Delta_{\ell_1 \ell_2 \ell_3}}
    \ B^{i j k, \ prim}_{\ell_1 \ell_2 \ell_3} \ (\mathcal{C}^{-1})^{i
      o}_{\ell_1} \ (\mathcal{C}^{-1})^{j p}_{\ell_2} \
    (\mathcal{C}^{-1})^{k q}_{\ell_3} \ B^{o p q, \ prim}_{\ell_1
      \ell_2 \ell_3 } \right]^{-1} \! \cdot \ \mathcal{S}_{prim} \, ,
\end{equation}
where $\Delta_{\ell_1 \ell_2 \ell_3} = 6$, when $\ell_1 = \ell_2 =
\ell_3$, 2, when $\ell_1 = \ell_2 \neq \ell_3$ or $\ell_1 \ne \ell_2 =
\ell_3$, and 1 otherwise. We further introduced the theoretical
bispectrum for $f_{\NL} = 1, \ B^{i j k, \, prim}_{\ell_1 \ell_2
  \ell_3}$, which is defined as
\begin{equation}
  B^{i j k, \ prim}_{\ell_1 \ell_2 \ell_3}= 2 \, I_{\ell_1 \ell_2 \ell_3}
  \int dr \, r^2 \ \lbrack \beta^i_{\ell_1}(r) \beta^{j}_{\ell_2}(r)
  \alpha^{k}_{\ell_3}(r) + \beta^{k}_{\ell_3}(r) \beta^{i}_{\ell_1}(r)
  \alpha^{j}_{\ell_2}(r) + \beta^{j}_{\ell_2}(r) \beta^{k}_{\ell_3}(r)
  \alpha^{i}_{\ell_1}(r) \rbrack \, ,
\end{equation}
where the prefactor is given by
\begin{equation}
  I_{\ell_1 \ell_2 \ell_3} = \sqrt{\frac{(2 \ell_1 + 1)(2 \ell_2 +
      1)(2 \ell_3 + 1)}{4 \pi}} \,
  \wthreej{\ell_1}{\ell_2}{\ell_3}{0}{0}{0} \, .
\end{equation}

\begin{figure*}
  \label{fig:fnl_hist}
  \centerline{\resizebox{\hsize}{!}{\includegraphics*{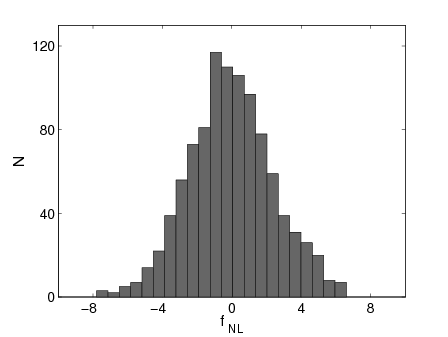} \,
      \includegraphics*{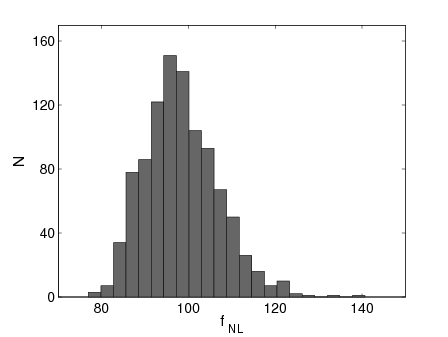}}}
  \caption{Histogram of the recovered \fnl\ values. We display the
    distribution of estimated \fnl\ values when applying the fast
    estimator to 1000 realization of temperature and polarization CMB
    maps. The input values used for the simulations were $f_{\NL} = 0$
    (\emph{left panel}), and $f_{\NL} = 100$ (\emph{right panel}).}
\end{figure*}

We used the equations above to implement the fast estimator for
temperature and polarization. As our primary goal is to validate our
simulation algorithm, we do not take into account possible
instrumental effects, sky cut, or noise. To test our simulations, we
generate two sets of 1000 CMB temperature and polarization maps with
resolution parameter $n_{side} = 512$, and $\ell_{max} = 1024$. We
consider one sample of purely \gn\ realizations of the CMB sky
($f_{\NL} = 0$), and one \ngn\ sample with a fiducial value of
$f_{\NL} = 100$. We then run the fast estimator on the maps to compute
an estimate of \fnl. We show the distribution of the derived values in
\fig{fig:fnl_hist}. We find the input parameters to be recovered well,
the means of the distributions are $\langle f^{\mathrm{G}}_{\NL}
\rangle = -0.1$ and $\langle f^{\mathrm{NG}}_{\NL} \rangle = 98.4$ for
the \gn\ and \ngn\ simulations, respectively. The estimated standard
deviations are $\sigma_{f^{\mathrm{G}}_{\NL}} = 2.4$ and
$\sigma_{f^{\mathrm{NG}}_{\NL}} = 8.4$, compared to the expected error
predicted from a Fisher information matrix analysis of
$\sigma^{Fisher}_{f_{\NL}} = 2.4$. We conclude that the algorithm
outlined in \sect{sec:basics} and implemented as described in
\sect{sec:code} produces valid realizations of \ngn\ CMB temperature
and polarization maps.

\section{Summary}
\label{sec:summary}

In this paper, we introduced a new algorithm to simulate temperature
and polarization CMB maps containing \ngy\ of arbitrary local type. In
the proposed scheme, we generate spherical harmonic coefficients of
the \gn\ potential as a function of conformal distance, taking into
account the proper radial correlations. Then, the potential is
transformed to pixel space to compute the associated \ngn\
contribution. Finally, we make use of the real space representation of
the transfer functions to perform the line of sight integral in order
to calculate \gn\ and \ngn\ contribution to the CMB maps.

We developed and applied a quadrature scheme that allows us to
increase the numerical efficiency of the code. As a starting point, we
derived an expression to quantitatively calculate the mean error
introduced by replacing the radial integral by a finite sum. On that
basis, we were able to choose both, nodes and weights for numerical
quadrature in an optimal way. As a last step, we successfully reduced
the memory consumption of the algorithm.

For {\emph WMAP5+BAO+SN} cosmological parameters, we simulated 1000
realizations of \ngn\ CMB temperature and polarization maps with
resolution parameters $n_{side} = 512$ and $\ell_{max} = 1024$. To
validate the algorithm, we applied the well studied and widely
accepted fast cubic (bispectrum) estimator to the simulations. For
both, a set of \gn\ and \ngn\ realizations of CMB sky maps, the input
parameters were consistently recovered. We make our simulations
publicly available at
\url{http://planck.mpa-garching.mpg.de/cmb/fnl-simulations}.

\acknowledgments
Some of the results in this paper have been derived using the HEALPix
\citep{2005ApJ...622..759G} package. BDW is partially supported by NSF
grants AST 0507676 and AST 07-08849. BDW gratefully acknowledges the
Alexander v. Humboldt Foundation's Friedrich Wilhelm Bessel Award. BDW
thanks the Caltech Astrophysics group for their hospitality while this
work was being completed.

\bibliographystyle{aa}
\bibliography{literature}

\end{document}